\documentstyle[12pt]{article}
\font\openbig=msym10  scaled\magstep1
\font\openscr=msym7 scaled\magstep1
\font\openscrscr=msym5 scaled\magstep1

\newfam\openfam
\textfont\openfam=\openbig
\scriptfont\openfam=\openscr
\scriptscriptfont\openfam=\openscrscr
\def\open{\fam\openfam}

\font\Scbig=cmss10  scaled\magstep1
\font\Scscr=cmss8 scaled\magstep1
\font\Scscrscr=cmss8

\newfam\Scfam
\textfont\Scfam=\Scbig
\scriptfont\Scfam=\Scscr
\scriptscriptfont\Scfam=\Scscrscr
\def\Sc{\fam\Scfam}

\makeatletter

\newdimen\normalarrayskip              
\newdimen\minarrayskip                 
\normalarrayskip\baselineskip
\minarrayskip\jot
\newif\ifold             \oldtrue            \def\new{\oldfalse}
\def\arraymode{\ifold\relax\else\displaystyle\fi} 
\def\@arrayskip{\ifold\baselineskip\z@\lineskip\z@
     \else
     \baselineskip\minarrayskip\lineskip2\minarrayskip\fi}
\def\@arrayclassz{\ifcase \@lastchclass \@acolampacol \or
\@ampacol \or \or \or \@addamp \or
   \@acolampacol \or \@firstampfalse \@acol \fi
\edef\@preamble{\@preamble
  \ifcase \@chnum
     \hfil$\relax\arraymode\@sharp$\hfil
     \or $\relax\arraymode\@sharp$\hfil
     \or \hfil$\relax\arraymode\@sharp$\fi}}
\def\@array[#1]#2{\setbox\@arstrutbox=\hbox{\vrule
     height\arraystretch \ht\strutbox
     depth\arraystretch \dp\strutbox
     width\z@}\@mkpream{#2}\edef\@preamble{\halign \noexpand\@halignto
\bgroup \tabskip\z@ \@arstrut \@preamble \tabskip\z@ \cr}%
\let\@startpbox\@@startpbox \let\@endpbox\@@endpbox
  \if #1t\vtop \else \if#1b\vbox \else \vcenter \fi\fi
  \bgroup \let\par\relax
  \let\@sharp##\let\protect\relax
  \@arrayskip\@preamble}

\makeatother

\begin{document}
\hfuzz=1pt

\date{(December 16, 1991)}
\title{\sc A Conformal Field Theory Formalism
from Integrable Hierarchies via the Kontsevich--Miwa Transform
\footnote{submitted for publication}}

        \author{{\large A.~M.~Semikhatov}\\
{\small {\sl Theory Division, P.~N.~Lebedev Physics Institute}}\\
{\small {\sl 53 Leninsky prosp., Moscow SU 117924 USSR}}}

\maketitle

\begin{abstract}
We attempt a direct derivation of a conformal field theory description of 2D
quantum gravity~+~matter from the formalism of integrable hierarchies subjected
to Virasoro constraints. The construction is based on a generalization of the
Kontsevich parametrization of the KP times by introducing Miwa parameters into
it. The resulting Kontsevich--Miwa transform can be applied to the Virasoro
constraints provided the Miwa parameters are related to the background charge
$Q$ of the Virasoro generators on the hierarchy. We then recover the field
content of the David-Distler-Kawai formalism, with the matter
theory represented by a scalar with the background charge
$Q_{\rm m}=Q-\frac{2}{Q}$. In particular, the
tau function is related to the correlator of a product of the `21' operators
of the minimal model with central charge $d=1-3Q^2_{\rm m}$.
\end{abstract}

\newpage

{\bf 1.~Introduction.}
The Matrix Models \cite{[BK],[DSh],[GM]}, besides their applications to
matter+gravity systems \cite{[DVV],[GGPZj],[DfK]}, have also
shown intriguing relations with integrable hierarchies subjected to Virasioro
constraints \cite{[D],[FKN1],[DVV],[MM],[IM]} as well as with the
intersection theory on the moduli space of curves \cite{[W1],[K],[W2]}.
However, a challenging problem remains of giving a direct proof of the
equivalence between
the `hierarchical' formalism and the conformal field theory
description of quantum gravity \cite{[KPZ],[DK],[Da]}. As a `direct proof' one
would like to have something more than just the circumstantial evidence. Thus,
it would be interesting to identify
the conformal field theory ingredients in the
machinery of integrable hierarchies. The only viable candidate for a
`space-time' of the saught conformal theory could be the spectral curve
associated to the hierarchy. This, however, is not enough: the infinite
collection of time parameters inherent to integrable hierarchies is hard to
deal
with in the standard conformal techniques.

Recall, however, that there does exist a formalism in which the time variables
are treated, in a sense, on equal footing with the spectral parameter. This is
the Miwa transform used in the KP hierarchy \cite{[Mi],[Sa]}. What is more, a
similar construction has been proposed recently by Kontsevich \cite{[K]}  and
used in \cite{[W2],[MMM]} in relation to the Virasoro constraints on integrable
hierarchies.

In this paper we show that what one needs in order to recast the Virasoro
constraints on the KP hierarchy into certain conformal
field theory data, is a generalization of the Kontsevich parametrization of
the KP times. That is, we view
Kontsevich' parametrization as
a special case of the Miwa's. However, the extra `degrees of freedom' present
in the Miwa transform are not completely
frozen: it turns out that one has to allow these to vary so as to be able to
move between {\it different} (generalised) Kontsevich transformations: we will
see that different Kontsevich transformations should be used depending on the
operators one considers. We thus get a `Miwa-parametrized set' of Kontsevich
transformations, which we call the Kontsevich--Miwa transform.

For each of the generalised Kontsevich transformations, pulling back the
Virasoro constraints to the Kontsevich parametrization results in relations,
analogous to the ``master equation'' of ref.\cite{[MS]} (see also
ref.\cite{[MMM]}), which happen to be nothing but equations on correlation
functions in an `auxiliary' conformal field theory, stating the decopling
\cite{[BPZ]} of a certain null vector.  This allows us to make contact with
refs.\cite{[DK],[Da]}. It is interesting to observe how the formalism of
quantum gravity+matter in the conformal gauge has its counterparts in an
interplay of discrete parameters inherent to the Miwa formalism.

{\bf 2.~Virasoro action on the KP hierarchy}. The KP hierarchy is described in
terms of $\psi {\rm Diff}$ operators \cite{[DDKM]} as an
infinite set of mutually commuting evolution equations
on the
coefficients $w_n(x,~t_1,~t_2,~t_3,\ldots)$ of a $\psi{\rm Diff}$
operator $K$ of the form (with $D = \partial /\partial x $)
\begin{equation}
K = 1 + \sum _{n\geq 1}w_n D^{-n}
\label{(1)}
\end{equation}
The wave function and the adjoint wave function are defined by
\begin{equation}
\psi (t,z) = Ke^{\xi (t,z)} ,\quad  \psi ^\ast (t,z) =
K^{\ast -1}e^{-\xi (t,z)},\quad
\xi (t,z) = \sum _{r\geq 1}t_r z^r
\label{(2)}
\end{equation}
where $K^\ast $ is the formal adjoint of $K$.
The wave functions are related to the tau function via
\begin{equation}
\psi(t,z) = e^{\xi (t,z)}{\tau (t-[z^{-1}])\over \tau (t)},\quad
\psi^\ast (t,z) =
e^{-\xi (t,z)}{\tau (t+[z^{-1}])\over \tau (t)}
\label{(3)}
\end{equation}
where
$
t \pm  [z^{-1}] = (t_1 \pm  z^{-1}, t_2 \pm  {1\over 2}
z^{-2}, t_3 \pm  {1 \over 3} z^{-3}, \ldots)$.

The Virasoro action on the tau function is implemented by
the generators,
\begin{equation}
\new
\begin{array}{rcl}
{\Sc L}_{p>0} &=& {1\over 2}\sum ^{p-1}_{k=1}{\partial^{2} \over
\partial t_{p-k} \partial t_k} + \sum _{k\geq 1}kt_k {\partial \over \partial
t_{p+k}} + (a_0 + (J - {1\over 2})p) {\partial \over \partial t_p}\\
{\Sc L}_0&=& \sum _{k\geq 1}kt_k {\partial \over \partial t_k} + {1\over
2}a^2_0 - {1\over 24} \\
{\Sc L}_{p<0} &=& \sum _{k\geq 1}(k - p)t_{k-p} {\partial \over \partial t_k}+
{1\over 2}\sum ^{-p-1}_{k=1}k(-p - k)t_kt_{-p-k}+ (a_0 + (J -
{1\over 2})p)(-p)t_{-p}
\end{array}
\label{Lontau}
\end{equation}
which satisfy the Virasoro algebra with central charge $-12
(J^2 - J + {1\over 6})$. It will be useful to introduce an
``energy-momentum tensor"
 ${\Sc T} (u) = \sum _{p\in  { \open Z}}u^{-p-2} {\Sc L}_p$.

{\bf 3.~Miwa--Kontsevich transform.} The Miwa reparametrization
of the KP times is accomplished by the substitution
\begin{equation}
t_r = {1\over r} \sum_j n_j z^{-r}_j
\label{Miwatransform}
\end{equation}
where $\{z_j \}$ is a set of points on the complex plane.
By the Kontsevich transform we understand the dependence, via
eq.(\ref{Miwatransform}), of $t_r$ on the $z_j$ for {\it fixed} $n_j$.

To recast the
Virasoro constraints ${\Sc L}_{n\geq -1}\tau =0$ into the Kontsevich
paremetrization, note that picking out the involved ${\Sc L}$'s amounts to
retaining in ${\Sc T}(z)$ only terms with $z$ to negative powers:
\begin{equation}
{\Sc T}^{(\infty )}(v) = \sum _{n\geq 0}v^{-n-1} {1\over 2\pi
i} \oint dz z^n {\Sc T}(z) = {1\over 2\pi i} \oint dz {1\over v - z}  {\Sc
T}(z)
\label{(8)}
\end{equation}
where $v$ is
from a neighbourhood of the infinity and the integration contour encompasses
this neighbourhood.

A crucial simplification is achieved by evaluating
$ {\Sc T}^{(\infty )}(v)$ only at a point from the above set $\{z_j \}$ (one
has to take care that it lie in the chosen neighbourhood).
It is straightforward to see that ${\Sc T}^{(\infty )}(z_i)$ in the
Kontsevich-Miwa parametrization is given by the operator
\begin{equation}
\new
\begin{array}{rcl}
{\cal T}_{\{n\}}(z_i) &=& {1\over 2\pi i}
\oint dz { 1\over z_i - z} \left\{ (J - {1\over 2} ) {1\over z}\sum _{r\geq
1}z^{-r-1} {\partial \over \partial t_r} + {1\over 2}\sum _{r,s}z^{-r-s-2}
{\partial ^2\over \partial t_r\partial t_s} \right.\\
{}&+& \left. \sum  _j n_j {1\over z_j -
z}\sum _{r\geq 1}z^{-r-1} {\partial \over \partial t_r} + {1\over 2}\sum_j {n_j
+ n_j^2\over (z_j - z)^2} + {1\over 2}\sum _{^{j,k}_{j\neq k}}{n_jn_k\over (z_j
- z)(z_k - z)} \right. \\
 {}&-& \left. J\sum_j {n_j\over (z_j - z)^2} + (J - {1\over 2}) \sum
_{r\geq 1}z^{-r-2} r {\partial \over \partial t_r} \right\}
 \end{array}
 \label{(9)}
 \end{equation}
which depends parametrically on the chosen $n_j$.
Our aim, however, is to express everything through the $\partial
/\partial z_j$ derivatives, as the tau function should be viewed in the
Kontsevich parametrization as a function of the $z_i$.
Unfortunately, one cannot substitute $\partial /\partial t_r$ in terms
of $\partial
/\partial z_j$, as the equation relating $t_r$ and $z_j$ does not allow this.
However, when we evaluate the residues in (\ref{(9)}) we find that
the $t$-derivatives arrange into the combinati\-ons which are
just the desired $\partial /\partial {z_j}$'s,
apart from
the term
\begin{equation}
- \left( J - {1\over 2} - {1\over 2n_i}\right) \sum _{r\geq 1}r
z^{-r-2}_i {\partial \over \partial t_r}
\end{equation}
which should thus be set to zero by
choosing
\begin{equation}
n_i = {1\over 2J - 1} \equiv  {1\over Q}
\label{ni}
\end{equation}
In this way we arrive at\footnote{A more detailed derivation, as well as an
extension to $N$-reduced KP hierarchies, is given in \cite{[S1]}.}
\begin{equation}
{\cal T}_{\{n\}}(z_i) =
-{Q^2\over 2} {\partial ^2\over \partial z_i^2 } - \sum _{j\neq i} {1\over z_j
-
z_i} \left( {\partial \over \partial z_j} -
Qn_j{\partial \over \partial z_i}\right)
\label{Tgen}
\end{equation}
This operator depends on the collection of the $n_j$ with $j\neq i$. These are
optional, and can be chosen freely. In particular, if one wishes {\it all} the
${\Sc T}^{(\infty )}(z_j)$ to carry over to the Kontsevich variables along with
${\Sc T}^{(\infty )}(z_i)$, all the
$n_j$ have to be fixed to the same value (\ref{ni}). Then, one gets
``symmetric'' operators
\begin{equation}
{\cal T}(z_i) =
-{Q^2\over 2} {\partial ^2\over \partial z_i^2 } - \sum _{j\neq i} {1\over z_j
-
z_i} \left( {\partial \over \partial z_j} - {\partial \over \partial
z_i}\right)
\label{T part}
\end{equation}
These, of course, satisfy the centreless algebra spanned by the $\{ {\Sc
L}_{n\ge -1}\}$ Virasoro generators. Then,
if one starts with the {\it Virasoro}-{\it constrained} KP
hierarchy, {\it i.e.},
${\Sc T}^{(\infty )}(z)\tau = 0,$
one ends up in the Kontsevich parametrization with the KP Virasoro {\sl master
equation} (cf. ref.\cite{[MS]})
$ {\cal T}(z_i).\tau \{z_j \} = 0$.

{\bf 4.} Now, consider the subject which is is apparently quite different from
what we had in the previous section. Introduce a
conformal theory of a $U(1)$ current $j(z)=
\sum_{n\in {\open Z}} j_n z^{-n-1}$ and an energy-momentum tensor $T(z)=
\sum_{n\in {\open Z}}L_n z^{-n-2}$:
\begin{equation}
\new \begin{array}{lrcl}
\left[ j_m ,\right.&\left.\!\!\!\! j_n \right] &=& km \delta_{m+n,0}\\
\left[ L_m ,\right.&\left. \!\!\!\!L_n \right] &=& (m-n)L_{m+n} + {d+1 \over
12}(m^3 - m)\delta_{m+n,0}\\ \left[ L_m ,\right.&\left. \!\!\!\!j_n \right] &=&
-nj_{m+n} \end{array}
\label{thetheory}
\end{equation}
(We have parametrized the central charge as $d+1$).  Let $\Psi$ be a primary
field with conformal dimension $\Delta$ and $U(1)$ charge $q$. Then, in the
standard setting of \cite{[BPZ]}, we find a null vector at level 2:
\begin{equation} | \Upsilon \rangle
= \left(\alpha L_{-1}^2 + L_{-2} + \beta j_{-2} + \epsilon j_{-1} L_{-1}
\right)
| \Psi \rangle \label{Upsilon} \end{equation}
where \begin{equation}
\alpha = {k\over 2q^2}~,\quad \beta = -{q\over k} - {1\over 2q}~,\quad \epsilon
= -{1\over q}~,\quad \Delta = -{q^2\over k} - {1\over 2} \label{parameters}
\end{equation} with $q$ given by,
\begin{equation} {q^2\over k} = {d-13 \pm \sqrt{(25-d)(1-d)} \over 24}
\label{q2} \end{equation}
and, accordingly, \begin{equation} \Delta = {1-d \mp \sqrt{(25-d)(1-d)} \over
24}~.  \label{Delta}
\end{equation}

Factoring out the state (\ref{Upsilon}) leads in the
usual manner to the equation
\begin{equation}
\left\{ {k\over 2q^2}{\partial^2 \over \partial  z^2 } - {1\over q} \sum_{j}
{1\over z_j - z} \left(q{\partial\over \partial z_j}
- q_j{\partial\over \partial z} \right) +{1\over q}\sum_{j}{q\Delta_j - q_j
\Delta \over (z_j - z)^2}\right\}
\langle \Psi (z) \Psi_1 (z_1) \ldots \Psi_n (z_n) \rangle = 0
\label{decouplinggen}
\end{equation}
where $\Psi_j$ are primaries of dimension $\Delta_j$ and $U(1)$ charge $q_j$.
In particular,
\begin{equation}
\left\{ {k\over 2q^2}{\partial^2 \over \partial  z_i^2 } + \sum_{j\neq i}
{1\over z_i - z_j} \left({\partial\over \partial z_j} - {\partial\over \partial
z_i} \right) \right\} \langle \Psi (z_1) \ldots \Psi (z_n) \rangle = 0
\label{(21)}
\end{equation}
This equation will be crucial for the comparison with the KP hierarchy in
Sect.~5.

Writing the Hilbert space as
(matter$)\otimes($current$)\equiv {\cal M}\otimes
{\cal C}$,
$|\Psi \rangle =|\psi \rangle\otimes |\tilde {\Psi}\rangle$,
we introduce the matter Virasoro generators $l_n$ by,
\begin{equation}
L_n = l_n + \tilde{L}_n \equiv l_n + {1\over 2k}\sum_{m\in {\open
Z}}:j_{n-m}j_m :  \label{(23)}
\end{equation}
They then have central charge $d$. It turns out that
\begin{equation}
|\Upsilon \rangle = \left({k\over
2q^2}l_{-1}^2 + l_{-2} \right) |\Psi \rangle
\label{(24)}
\end{equation}
and thus we are left with a null vector
in the matter Hilbert space ${\cal M}$. Now, the dimension of $|\psi \rangle$
in the matter sector,
\begin{equation}
\delta = \Delta - {1\over 2k}q^2 = {5-d\mp\sqrt{(25-d)(1-d)}\over 16},
\label{delta}
\end{equation}
is of course that of the `21' operator of the minimal model with central charge
$d$.

{\bf 5.~Conformal field theory from Virasoro constraints.}
A contact between sections 4 and 3, i.e., between
conformal field theory formalism
and the KP hierarchy is established by
assuming the ansatz
\begin{equation}
\tau \{z_j \}= \lim_{ n\rightarrow \infty} \langle \Psi (z_1)
\ldots \Psi (z_n)\rangle
\label{(13)}
\end{equation}
then, comparing eqs.(\ref{decouplinggen}) and (\ref{Tgen}), one finds
\begin{equation}
Q^2 = -{k\over q^2} = {13-d \pm
\sqrt{(25-d)(1-d)}\over 6},
\label{Q2}
\end{equation}
and $d$ is therefore
determined in terms of the parameter $Q$ from (\ref{Lontau})
(where $J={Q+1\over 2}$), as
\begin{equation}
d = 13 -3Q^2 - {12\over Q^2},
\label{d(Q)}
\end{equation}
Note that the energy-momentum tensor $T(z)$ appears to
have a priori nothing to do with the energy-momentum tensor on the hierarchy we
have started with. In terms of the latter tensor, eq.(\ref{Tgen})
comprises the contribution of all the positive-moded
Virasoro generators, while out of $T(z)$
only $L_{-1}$ and $L_{-2}$ enter the equivalent equation (\ref{Upsilon}).

To reconstruct matter theory field operators, consider the form the
${\Sc L}_{n\geq -1}$-Virasoro constraints take
for the wave function of the hierarchy, $w(t,z_k)\equiv e^{-\xi
(t,z_k)}\psi(t,z_k)$, which should now become a function of the $z_j$,
$w\{z_j\}(z_k)$.  More precisely, consider
the `unnormalized' wave function $ \bar{w}\{z_j\}(z_k) =
\tau\{z_j\}w\{z_j\}(z_k) $.  Then
the use of the Kontsevich transform at
a Miwa point
$n_j = 1/Q$, $j\neq k$ and $n_k = -1$,
gives\footnote{To obtain the insertion into the correlation function
(\ref{barw}) at the point $z_k$ of the operator we are interested in by itself,
rather
than its fusion with the `background' $\Psi$, we use the Kontsevich
transform at the value of the Miwa parameter $n_k=-1$ instead of ${1\over
Q}-1$.
This means that we are in fact considering $\bar{w}\{z_j\}^{}_{j\neq k}(z_k)$.
Similar remarks apply to the other correlation functions considered below.}
\begin{equation} \bar{w}\{z_j\}(z_k)=
\left\langle \prod_{j\neq k}
\Psi(z_j)\cdot\Xi(z_k)\right\rangle
\label{barw}
\end{equation}
where $\Xi$ is a primary field with the $U(1)$ charge $-qQ$ and dimension
$-Q\Delta$. Now we choose
in (\ref{Q2}) the branch of the square root $\sqrt{Q^2}$ so that
\begin{equation}
Q=-{1\over 2}\sqrt{{25 - d\over 3}} \mp {1\over 2}\sqrt{{1 - d\over 3}}
\equiv {-Q_{\rm L} \mp Q_{\rm m} \over2}\equiv \alpha_\mp
\end{equation}
(with the upper/lower signs corresponding to those in (\ref{Q2})).
This establishes the
physical meaning of the background charge $Q$ present
initially in the Virasoro constraints. (Note that  it
has entered explicitly in the
Kontsevich transform through (\ref{ni}).)
Now, the dimension of $\Xi$ is equal to $\mp {1\over 2}Q_{\rm m}$, which
implies in turn that its  dimension in the matter sector
equals
\begin{equation}
\mp {1\over 2}Q_{\rm m} - {1\over
2k}\left(qQ\right)^2 = \mp{1\over 2}Q_{\rm m} + {1\over 2} \equiv
\left\{
\begin{array}{l}
1-J_{\rm m}\\
J_{\rm m}
\end{array}
\right.
\end{equation}
where $J_{\rm m}$ is the conformal `spin' (dimension) of a  $bc$ system.
Thus, although in an indirect way, the wave function is
associated
(for, say, the lower signs) with the $b$-field $B$ of a $bc$
system\footnote{Note that this is an {\it unconstrained}
$bc$ system, i.e. {\it not}
the spin-$J$ one underlying the constraints ${\Sc L}_{n\geq -1}\tau = 0$.}.
The adjoint wave function is then
similarly related to the corresponding
$c$ field $C$: for instance, the function
$\tau (t - [z^{-1}_k] + [z^{-1}_l])$ is annihilated by the
operator
\begin{equation}
-{Q^2\over 2} {\partial ^2\over \partial z_i^2 } -
\sum_{{j\neq i,~j\neq k}{\atop{j\neq l}}}{1\over z_j - z_i}
\left( {\partial \over \partial z_j} -
{\partial \over \partial z_i}\right) + Q\left({1\over z_l - z_i}-
{1\over z_k - z_i}\right){\partial\over \partial z_i}
\end{equation}
and thus is proportional to the correlation function
\begin{equation}
\left\langle \prod_{{j\neq k}\atop{j\neq l}}\Psi(z_j)
e^{{qQ\over k}\int^{z_k}j}B(z_k)
e^{-{qQ\over k}\int^{z_l}j}C(z_l)\right\rangle =
\left\langle \prod_{{j\neq k}\atop{j\neq l}} \Psi(z_j)
(z_k - z_l)
e^{{qQ\over k}\int^{z_k}_{z_l}j}B(z_k) C(z_l)\right\rangle
\label{BC}
\end{equation}

Thus the whole `Borel' subalgebra of the $W_\infty (J)$ algebra \cite{[BVdW]},
which is the symetry algebra of the Virasoro-constrained KP hierarchy
\cite{[S4]}, is represented in terms of the bilocal operator insertions, read
off from (\ref{BC}), placed at points from the {\it fixed set}
$\{z_j\} \! \times \! \{z_{j^{\prime}}\} \! \subset\! {\open CP}^1\!\times\!
{\open CP}^1 $. (Thus, although it is tempting
to take in (\ref{BC}) the limit $z_k
\rightarrow z_l$, this cannot be done naively, as it would affect the whole
construction of the Kontsevich-Miwa transform~!)

As a cross-check, it is interesting to compare (\ref{BC}) with a well-known
formula valid for a general (i.e., not necessarily Virasoro-constrained) KP
tau-function:
\begin{equation}
{\tau (t - [u^{-1}] + [z^{-1}])\over \tau (t)}=
(u-z) e^{\xi (t,z)- \xi (t,u)}
\partial^{-1}
\left(\psi (t,u)\psi^\ast (t,z)\right)
\label{tauovertau}
\end{equation}
In the Kontsevich parametrization,
\begin{equation}
e^{\xi (t,z)- \xi (t,u)}=\prod_j\left({z_j-u\over z_j-z}\right)^{n_j}
\end{equation}
On the other hand, fusing the exponential factor in (\ref{BC}) with the product
of the $\Psi(z_j)$, gives the factor
\begin{equation}
\prod_{{j\neq k}\atop{j\neq l}}\left({z_j-z_k\over z_j-z_l}\right)^{{q\over
k}\left(-{qQ\over k}\right)\cdot k} =
\prod_{{j\neq k}\atop{j\neq l}}\left({z_j-z_k\over z_j-z_l}\right)^{1\over Q}
\end{equation}
which is in perfect agreement with (\ref{ni}).

The $BC$
system has
central charge $1-3Q^2_{\rm m}=d$.
By bosonization one gets a scalar $\varphi$ with the energy-momentum tensor
\begin{equation}
T_{\rm m} = -{1\over 2}\partial \varphi \partial \varphi
+ {i\over 2}Q_{\rm m}\partial^2\varphi,
\label{Tmatter}
\end{equation}
thus establishing the relation with
minimal models \cite{[BPZ],[DF],[FQS]} (for appropriate values of $d$).

Further, as to
the theory in ${\cal C}$, recall that we have
\begin{equation}
[j_m,j_n]=km\delta_{m+n,0},\quad j_{n>0}|\Psi\rangle =
0, \quad j_0|q\rangle = q|\Psi\rangle
\end{equation}
with
negative $q^2/k$\ \footnote{for $d<1$. For $d>25$, on the other hand, $q^2/k$
is
positive, but then one has to consider the hierarchy with {\it imaginary}
$Q$~! It appears that the matter and the Liouville theory then take place of
one another, and $Q=i\alpha$ with $\alpha$ being the cosmological constant.}.
To see what the current corresponds to in the KP theory, consider
the correlation function with an extra insertion of an operator which depends
on
only $j$:
\begin{equation}
\left\langle \prod_{j\neq k\atop j\neq l} \Psi(z_j)
e^{\omega\int^{z_k}_{z_l}j}\right\rangle
\label{expj}
\end{equation}
The operator inserted in the background of the $\Psi$'s has $U(1)$ charge
$k\omega$ (with respect to $z_k$). Now, for (\ref{expj}) to come from the
KP hierarchy, $\omega$ must be equal to
$2\Delta /q$. Then the decoupling equation states that the correlation function
(\ref{expj}) is annihilated by the operator
\begin{equation}
{\cal T}(z_i) + (Q^2 - 2)\left({1\over z_k - z_i} -  {1\over z_l - z_i}\right)
{\partial\over \partial z_i}
\end{equation}
and therefore coincides, up to a constant, with the tau function $\tau (t)$
evaluated at the Miwa point
\begin{equation}
n_j = \left\{
\begin{array}{ll}
{1\over Q}, &j\neq k,\quad j\neq l\\
Q-{2\over Q}, &j=k\\
-\left(Q-{2\over Q}\right), &j=l
\end{array}
\right.
\end{equation}
Her, $Q-{2\over Q}=\mp Q_{\rm m}$; abusing the notations, the function we
are considering can be written as $\tau\{z_j\}^{\phantom{W}}_{{j\neq k}\atop
{j\neq l}}(t-Q_{\rm m}[z_k^{-1}]+ Q_{\rm m}[z_l^{-1}])$.  This is another
illustration of how the Kontsevich-Miwa transform works: establishing the
relation to the conformal field theory of different operators requires fixing
different values of the $n_j$.

The balance of
dimensions and $U(1)$ charges of both the $\Psi$ and $\Xi$ operators follows a
particular pattern. That is, as there are no $1/(z_i-z_j)^2$-terms in the
master equation, we can only derive from it operators from a special sector,
i.e., those whose dimensions
$\Delta_j$
and $U(1)$ charges $q_j$ satisfy (see (\ref{decouplinggen}),
\begin{equation}
\Delta_j = \Delta{q_j\over q}
\label{Deltaj}
\end{equation}
(we continue to denote by $\Delta$ and $q$ {\it the} dimension and $U(1)$
charge
from (\ref{thetheory}) -- (\ref{Delta}), i.e., those of $\Psi$.) Then, the
dimension in the matter sector ${\cal M}$ is equal to
\begin{equation}
\delta_j = \Delta_j -{q_j^2\over 2k} = \Delta{q_j\over q} - {q_j^2\over 2k}
\label{deltaj}
\end{equation}
As the coefficient at the term linear in $q_j/\sqrt{-k}$ is $\frac{1}{2}Q_{\rm
m}$, this equation will always be satisfied for the matter operators
$e^{i\gamma\varphi}$ provided $q_j/\sqrt{-k}=\gamma$~! -- Thus
the recipe for a `dressing' inherited from the KP hierarchy states that
the two scalars $\varphi$ and $\phi$  enter the exponents with the same
coefficient.
Therefore, although
the field content is the same as in ref.\cite{[DK]},
the David-Distler-Kawai formalism is not recovered directly from the KP
hierarchy\footnote{This can be seen also by noticing that the dimensions in
${\cal M}$ and ${\cal C}$ do not add up to 1; nor is the central charge equal
to
26. This is not a surprise, since the current $j$ is not anomalous.}.
Our \begin{equation}
\frac{q_j}{\sqrt{-k}}=-\frac{\sqrt{-k}\Delta}{q} \pm \frac{1}
{2\sqrt3}\sqrt{1-d+24\delta_j}
\end{equation}
differs from eq.(3.12) of \cite{[DK]}
\begin{equation}
\beta_j = -\frac{1}{2}Q_{\rm L}\pm \frac{1}
{2\sqrt3}\sqrt{1-d+24\delta_j}
\end{equation}
by the cosmological constant $\alpha=Q$.

The `bulk' dimensions $\Delta_j$, rather than being equal to 1, are related to
the gravitational scaling dimensions of fields. Indeed, evaluating the
gravitational scaling dimension of $\psi$ according to \cite{[Da],[DK],[KPZ]},
\begin{equation}
\hat {\delta}_{\pm} = {\pm\sqrt{1-d+24\delta}-\sqrt{1-d} \over
\sqrt{25-d}-\sqrt{1-d}}
\end{equation}
one would find
\begin{equation}
\hat{\delta}_+ = {3\over 8}
\pm {d-4-\sqrt{(1-d)(25-d)}\over 24} \label{hat delta}
\end{equation}
with the sign on the RHS corresponding to that in (\ref{q2})
and the subsequent formulae. In particular, choosing the {\it lower} signs
throughout, we have $\hat{\delta}_+ = \Delta + {1\over 2}$. More generally,
the gravitational scaling dimensions corresponding to (\ref{deltaj}) equal
\begin{equation}
\hat {\delta}_{j+} = -{q_j q\over k}= \Delta_j + {1\over
2}{q_j\over q} =\Delta_j + {1\over 2}Q{q_j\over \sqrt{-k}}
\end{equation}
(Again, this is valid for the `+'-gravitational scaling dimensions {\it and}
the
lower signs in eqs.(\ref{Q2}) etc., i.e., for only one out of four
possibilities
to choose the signs.)

{\bf 5.~Concluding remarks.}~1.~
Various aspects of the conversion of Virasoro constraints into
decoupling equations deserve more study from the `Liouville'
point of view. The Kontsevich-type matrix integral whose
Ward identities coincide with our master equation may thus
provide a candidate for a
discretized definition of the Liouville theory.

2.~It was implicitly understood in the above that the matter
central charge
$d$ should be fixed to the minimal-models series; then factoring out the
null-vector leads to a bona fide minimal model. Now, thinking in terms of the
minimal models, how can the
{\it higher} null-vectors be arrived at starting from the Virasoro-constrained
hierarchies?

3.~It is interesting to study the region $1<d<25$ where $Q$ is complex,
in terms
of the Virasoro-constrained KP hierarchy; the old struggle with $1<d<25$ may be
carried on with a new hope in the realm of integrable hierarchies.

4.~It may be possible to extend the above Kontsevich-Miwa transform
`off-shell',
i.e., off the Virasoro constraints. Both of the two classes of objects, the tau
function etc., and the theory in ${\cal M}\otimes{\cal C}$, exist by
themselves,
while we have seen that imposing the Virasoro constraints on the one end
results
in choosing an irreducible representation on the other; see
eq.(\ref{tauovertau}) and the accompanying remark.

              \end{document}